\documentclass[apjl]{emulateapj}
\usepackage{mathptmx}

\newcommand{\ltsim}{\raisebox{-.5ex}{$\;\stackrel{<}{\sim}\;$}}
\newcommand{\gtsim}{\raisebox{-.5ex}{$\;\stackrel{>}{\sim}\;$}}
\newcommand{\kms}{\ifmmode {\rm km\ s}^{-1} \else km s$^{-1}$\fi}
\newcommand{\vFWHM}{\ifmmode V_{\mbox{\tiny FWHM}} \else
            $V_{\mbox{\tiny FWHM}}$\fi}
\newcommand{\msun}{$M_{\odot}$}
\newcommand{\et}{et al.\ }

\newcommand{\hb}{H$\beta$}
\newcommand{\mbh}{$M_{\rm BH}$}
\newcommand{\lledd}{$L/L_{\rm Edd}$}
\newcommand{\lbol}{$L_{\rm bol}$}
\newcommand{\aox}{$\alpha_{\rm ox}$}
\newcommand{\nh}{$N_{\rm H}$}
\newcommand{\Ka}{\hbox{Fe K$\alpha$}}
\newcommand{\xmm}{{\hbox{\sl XMM-Newton}}}
\newcommand{\chandra}{{\sl Chandra}}
\newcommand{\xray}{\hbox{X-ray}}

\journalinfo{The Astrophysical Journal, 682:???--???, 2008 July 20}
\slugcomment{Received 2007 September 24; accepted 2008 April 03}

\shorttitle{HARD X-RAY SPECTRUM AS A PROBE TO BH GROWTH}
\shortauthors{SHEMMER ET AL.}

\begin{document}

\title{The Hard X-ray Spectrum as a Probe for Black-Hole Growth in
  Radio-Quiet \\ Active Galactic Nuclei}

\author{
Ohad Shemmer,\altaffilmark{1}
W.~N. Brandt,\altaffilmark{1}
Hagai Netzer,\altaffilmark{2}
Roberto Maiolino,\altaffilmark{3}
and Shai Kaspi\altaffilmark{2,4}
}

\altaffiltext{1} {Department of Astronomy \& Astrophysics,
  Pennsylvania State University, University Park, PA 16802, USA;
  ohad@astro.psu.edu.}

\altaffiltext{2} {School of Physics \& Astronomy, Raymond and Beverly
  Sackler Faculty of Exact Sciences, Tel Aviv University, Tel Aviv
  69978, Israel.}

\altaffiltext{3} {INAF - Osservatorio Astronomico di Roma, via di
  Frascati 33, 00040 Monte Porzio Catone, Italy.}

\altaffiltext{4} {Physics Department, Technion, Haifa 32000, Israel.}

\begin{abstract}
  We study the hard-\xray\ spectral properties of ten highly luminous
  radio-quiet (RQ) active galactic nuclei (AGNs) at $z=1.3-3.2$,
  including new \xmm\ observations of four of these sources. We find a
  significant correlation between the normalized accretion rate
  (\lledd) and the hard-\xray\ photon index ($\Gamma$) for 35
  moderate--high luminosity RQ AGNs including our ten highly luminous
  sources. Within the limits of our sample, we show that a measurement
  of $\Gamma$ and $L_{\rm X}$ can provide an estimate of \lledd\ and
  black-hole (BH) mass (\mbh) with a mean uncertainty of a factor of
  \ltsim3 on the predicted values of these properties. This may
  provide a useful probe for tracing the history of BH growth in the
  Universe, utilizing samples of \xray-selected AGNs for which \lledd\
  and \mbh\ have not yet been determined systematically. It may prove
  to be a useful way to probe BH growth in distant Compton-thin type~2
  AGNs. We also find that the optical--\xray\ spectral slope (\aox)
  depends primarily on optical--UV luminosity rather than on \lledd\
  in a sample of RQ AGNs spanning five orders of magnitude in
  luminosity and over two orders of magnitude in \lledd. We detect a
  significant Compton-reflection continuum in two of our highly
  luminous sources, and in the stacked \xray\ spectrum of seven other
  sources with similar luminosities, we obtain a mean relative Compton
  reflection of $R=0.9^{+0.6}_{-0.5}$ and an upper limit on the
  rest-frame equivalent width of a neutral \Ka\ line of 105\,eV. We do
  not detect a significant steepening of the \xray\ power-law spectrum
  below rest-frame 2\,keV in any of our highly luminous sources,
  suggesting that a soft-excess feature, commonly observed in local
  AGNs, either does not depend strongly on \lledd, or is not
  accessible at high redshifts using current \xray\ detectors. None of
  our highly luminous sources displays \xray\ flux variations on
  timescales of $\sim$1\,hr, supporting the idea that the timescale of
  \xray\ variability depends inversely on \mbh\ and does not depend on
  \lledd.
\end{abstract}


\section{Introduction}
\label{sec_introduction}

The \xray\ spectrum of an unobscured, radio-quiet (RQ) AGN in the
\hbox{$\sim$2--100\,keV} energy band is best characterized by a single
power-law continuum of the form $N(E)\propto E^{-\Gamma}$, where
$\Gamma$, hereafter the photon index in the \hbox{$\sim$2--100\,keV}
energy band, typically lies in the range $\sim$1.5--2.5. A corona of
hot electrons is assumed to produce the hard-\xray\ emission via
Compton upscattering of UV--soft-\xray\ photons from the accretion
disk, and $\Gamma$ is predicted to be only weakly sensitive to large
changes in the electron temperature and the optical depth in the
corona (e.g., Haardt \& Maraschi 1991; Zdziarski \et 2000; Kawaguchi
\et 2001).
The relatively narrow range of $\Gamma$ values in RQ AGNs has been
reported in numerous studies (e.g., Nandra \& Pounds 1994, Reeves \&
Turner 2000; Page \et 2005; Shemmer \et 2005; Vignali \et 2005; Just
\et 2007), and typically no strong dependence of $\Gamma$ on redshift
or luminosity has been detected across the widest possible ranges of
these parameters.
On the other hand, a strong anticorrelation between $\Gamma$ and the
full width at half-maximum intensity (FWHM) of the broad emission-line
region (BELR) component of \hb\ has been found, first by Brandt \et
(1997).

The remarkable dependence between \xray\ and optical spectroscopic
properties has been suggested to arise from a more fundamental
correlation between $\Gamma$ and the accretion rate (e.g., Brandt \&
Boller 1998; Laor 2000). A high accretion rate is expected to soften
(steepen) the hard-\xray\ spectrum by increasing the level of disk
emission, resulting in the production of softer photons, which
increase the Compton cooling of the corona.
Using recent scaling relations for the BELR size, luminosity, and the
width of the broad \hb\ emission line from reverberation-mapping
studies, it is clear that the normalized accretion rate (i.e.,
\lbol/$L_{\rm Edd}$, hereafter \lledd, where \lbol\ is the bolometric
luminosity) is proportional to FWHM(\hb)$^{-2}$, at least for
low--moderate luminosity AGNs in the local universe (e.g., Kaspi \et
2000).

Subsequent \xray\ studies of nearby ($z$\ltsim0.5) unobscured RQ AGNs
have confirmed the Brandt \et (1997) \hbox{$\Gamma$-FWHM(\hb)}
anticorrelation (e.g., Leighly 1999; Reeves \& Turner 2000; Porquet
\et 2004; Piconcelli \et 2005; Brocksopp \et 2006), and others have
found significant correlations between $\Gamma$ and \lledd\ (e.g., Lu
\& Yu 1999; Porquet \et 2004; Wang \et 2004; Bian 2005). However, all
these studies were not able to disentangle the strong
\hbox{FWHM(\hb)-\lledd} dependence. Recently, Shemmer \et (2006;
hereafter S06) have suggested that this degeneracy can be removed if
highly luminous sources are included in the analysis. This can be
achieved by obtaining high-quality near-IR spectroscopy of the \hb\
spectral region, to obtain \lledd\ (e.g., Shemmer \et 2004), as well
as accurate measurements of $\Gamma$ using \xmm\ and \chandra\ for
highly luminous AGNs found at 1\ltsim$z$\ltsim3.

Within the limits of their sample of 30 sources, spanning three orders
of magnitude in luminosity, S06 have shown that $\Gamma$ does not
depend on luminosity or black-hole (BH) mass (\mbh). They have also
shown that the $\Gamma$ values of the five highly luminous sources in
their sample, while consistent with the values expected from their
normalized accretion rates (\lledd), are significantly higher than
expected from the widths of their broad \hb\ emission lines. This has
enabled, for the first time, breaking of the FWHM(\hb)-\lledd\
degeneracy and has provided evidence that $\Gamma$ depends primarily
on the accretion rate. However, the number of highly luminous sources
was small, and as explained below, some uncertainties remained.

In this work, we double the number of highly luminous sources at high
redshift and reinforce the S06 results. We show that $\Gamma$ can be
considered a reliable accretion-rate indicator for moderate--high
luminosity RQ AGNs, and that the combination of $\Gamma$ and \xray\
luminosity may provide a useful probe for tracing the history of BH
growth in the Universe. We also discuss the \xray\ spectral and
temporal properties of luminous, high-accretion rate AGNs at high
redshift as well as the dependence of the optical--\xray\ spectral
energy distribution (SED) on luminosity and \lledd.
In \S\,\ref{sec_observations} we describe our sample selection, and
present the new observations and their analysis. Our results are
presented and discussed in \S\,\ref{sec_results}, where we focus on
the correlation between the hard-\xray\ photon index and the
normalized accretion rate in RQ AGNs and its implications for probing
BH growth in the Universe. A summary of our main findings is given in
\S\,\ref{sec_conclusions}.
Throughout this work we consider only RQ AGNs to avoid any
contribution from jet-related emission to the \xray\ spectra.
Luminosity distances are computed using the standard cosmological
model with $\Omega_{\Lambda}=0.7$, $\Omega_{M}=0.3$, and
$H_{0}=70$\,\kms\,Mpc$^{-1}$.

\section{Sample Selection, Observations, and Data Analysis}
\label{sec_observations}

\subsection{Sample Selection}
\label{sec_sample}

\begin{figure}
\centering
\plotone{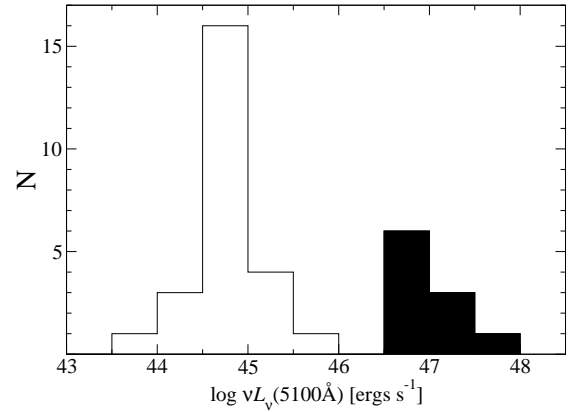}
\caption{Luminosity histogram of our core sample of 35 AGNs. The
  shaded region marks the ten highly luminous sources at $z=1.3-3.2$
  (the 25 moderate-luminosity sources are at $z<0.5$).}
\label{fig_hist}
\end{figure}

Our sample is composed of the 30 sources studied in S06 as well as
five highly luminous sources ($46\ltsim \log [\nu L_{\nu}
(5100~\mbox{\AA})] \ltsim 48$) whose hard-\xray\ spectra are analyzed
in this work for the first time. Motivated by the S06 hypothesis that
only highly luminous AGNs may break the \hbox{FWHM(\hb)-\lledd}
degeneracy, we have primarily searched for highly luminous AGNs that
have high-quality \hb\ and hard-\xray\ spectroscopy. Our search of the
literature and archive yielded ten such sources, five which were
studied in S06 and five that are analyzed below. Our new core sample
of 35 sources includes unabsorbed, type~1 RQ AGNs with $44\ltsim \log
[\nu L_{\nu} (5100~\mbox{\AA})] \ltsim 48$ (i.e., a moderate--high
luminosity range; see Fig.~\ref{fig_hist}). Rest-frame optical data
for this sample was obtained from Neugebauer \et (1987), Boroson \&
Green (1992), Nishihara \et (1997), McIntosh \et (1999), Shemmer \et
(2004), and Sulentic \et (2006). Hard-\xray\ data were obtained from
Page \et (2004b), Piconcelli \et (2005), S06, and from this work (see
below). All of our sources were selected to have high-quality \xmm\
hard-\xray\ spectra (except for HE~0926$-$0201 that was
serendipitously observed with \chandra). We have intentionally limited
our sample to the moderate--high luminosity range in order to minimize
potential effects of optical and \xray\ variability (since the optical
and \xray\ data are not contemporaneous) as well as potential spectral
complexities due to Compton reflection, as discussed further below
(see also S06 for more details). In light of our selection criteria,
we caution that our sample is neither complete nor fully
representative of the AGN population as a whole; our main results may
therefore be subject to selection biases over the wide range of AGN
properties, which were not fully explored in this work.

\begin{deluxetable*}{lcccccccc}
\tablecolumns{9}
\tabletypesize{\scriptsize}
\tablewidth{0pc}
\tablecaption{{\sl XMM-Newton} Observation Log \label{tab_obs_log}}
\tablehead
{
\colhead{} &
\colhead{} &
\colhead{} &
\colhead{} &
\colhead{} &
\colhead{{\sc Observation}} &
\multicolumn{3}{c}{{\sc Net Exposure Time (ks) / Source Counts}} \\
\colhead{{\sc AGN}} &
\colhead{{\sc RA (J2000.0)}} &
\colhead{{\sc DEC (J2000.0)}} &
\colhead{$z$\tablenotemark{a}} &
\colhead{$N_{\rm H}$\tablenotemark{b}} &
\colhead{{\sc Start Date}} &
\colhead{MOS1} &
\colhead{MOS2} &
\colhead{{\sc pn}}
}
\startdata
LBQS\,0109$+$0213 & 01 12 16.91 & $+$02 29 47.6 & 2.349 &
3.25 & 2007 Jan 8 & 33.7 / 211 & 33.8 / 218 & 26.3 / 554 \\
2QZ\,J023805.8$-$274337 & 02 38 05.80 & $-$27 43 37.0 & 2.471 &
1.73 & 2006 Dec 22 & 34.6 / 206 & 34.0 / 183 & 23.9 / 586 \\
Q\,1318$-$113 & 13 21 09.38 & $-$11 39 32.0 & 2.306 & 2.80 & 2006 Dec
28 & 47.7 / 853 & 46.9 / 876 & 33.2 /
2621 \\
SBS\,1425$+$606 & 14 26 56.18 & $+$60 25 50.9 & 3.202 &
1.58 & 2006 Nov 12 & 20.6 / 232 & 20.9 / 255 & 7.0 / 304
\enddata
\tablecomments{The net exposure time represents the {\sc livetime}
  following the removal of flaring periods. The net source counts are
  in the 0.2--10.0~keV band.}
\tablenotetext{a}{Systemic redshift measured from the optical emission
  lines and obtained from Shemmer \et (2004).}
\tablenotetext{b}{Neutral Galactic absorption column density in units
  of $10^{20}$\,cm$^{-2}$ obtained from Dickey \& Lockman (1990).}
\end{deluxetable*}

\subsection{New \xmm\ Observations}
\label{sec_xmm}

\begin{figure*}
\centering
\epsscale{0.7}
\plotone{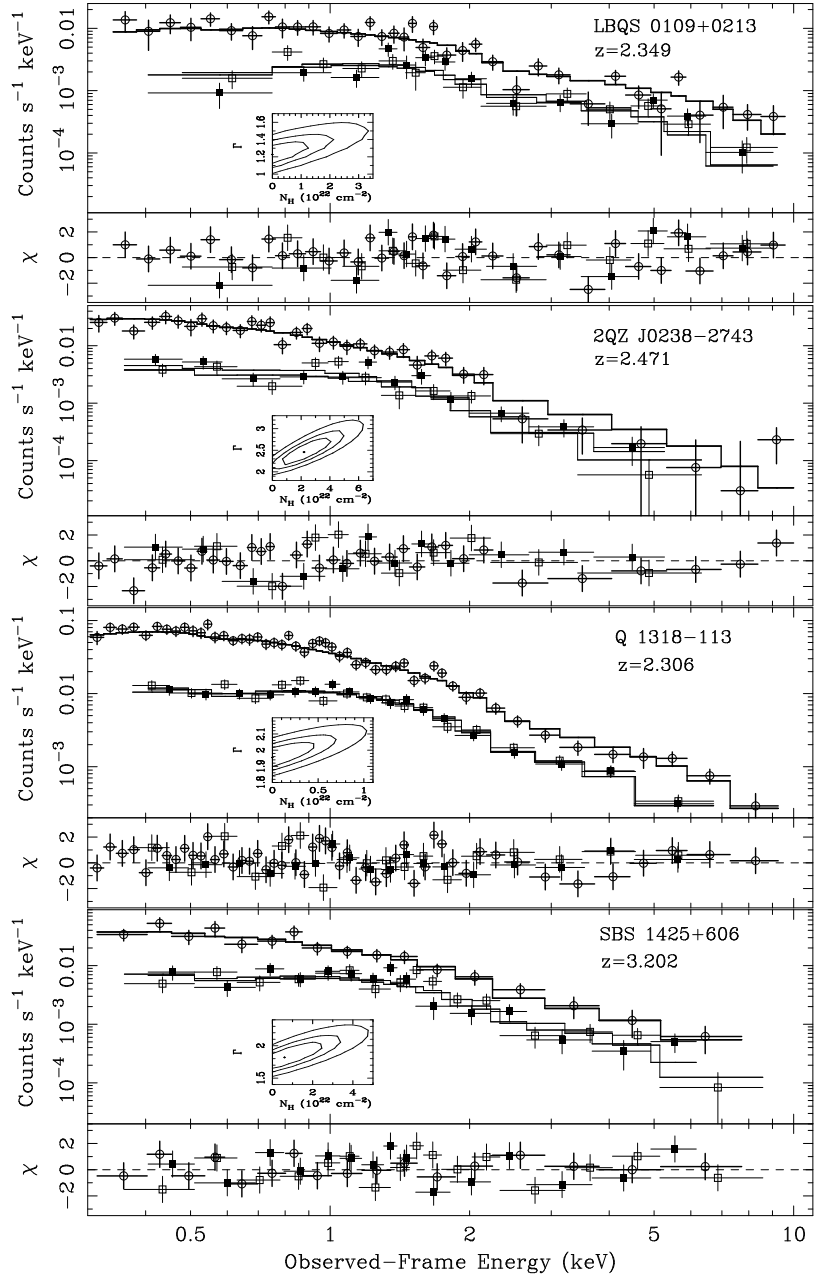}
\caption{Data, best-fit spectra, and residuals of our new \xmm\
  observations. Open circles, filled squares, and open squares
  represent the EPIC pn, MOS1, and MOS2 data, respectively. Solid
  lines represent the best-fit model for each spectrum, and the thick
  line marks the best-fit model for the pn data. The data were fitted
  with a Galactic-absorbed power-law model above a rest-frame energy
  of $\sim$2\,keV, and then extrapolated to 0.3\,keV in the observed
  frame. The $\chi$ residuals are in units of $\sigma$ with error bars
  of size 1. The insets show 68\%, 90\%, and 99\% confidence contours
  for $\Gamma$ and \nh, when the data are fitted with an additional
  neutral intrinsic-absorption component.}
\label{fig_spectra}
\end{figure*}

We have performed \xray\ spectral imaging observations of four new
sources from the Shemmer \et (2004) sample with \xmm\ (Jansen \et
2001); a log of these observations is presented in
Table\,\ref{tab_obs_log}. These sources were selected for \xmm\
observations for being luminous, high-accretion rate sources,
predicted to have high \xray\ fluxes, and having relatively low
Galactic column densities.
The data were processed using standard \xmm\ Science Analysis
System\footnote{http://xmm.esac.esa.int/sas} v6.5.0 tasks. The event
files of all the observations were filtered to remove periods of
flaring activity in which the count rates of each MOS (pn) detector
exceeded 0.35 (1.0) counts~s$^{-1}$ for events having $E>10$\,keV.
The time lost due to flaring in each observation varied between
\hbox{1\%--70\%} of the entire observing time; the net exposure times
in Table\,\ref{tab_obs_log} reflect the filtered data. The \xray\
spectra of the quasars were extracted from the images of all three
European Photon Imaging Camera (EPIC) detectors using apertures with
radii of 30\arcsec. Local background regions were at least as large as
the source regions. The spectra were grouped with a minimum of 20
counts per bin, except for the spectrum of Q~1318$-$113 which was
grouped with a minimum of 50 counts per bin. Joint spectral fitting of
the data from all three EPIC detectors for each source was performed
with {\sc xspec} v11.3.2 (Arnaud 1996). We employed Galactic-absorbed
power-law models at rest-frame energies \gtsim2\,keV, corresponding to
\gtsim0.5--0.6\,keV in the observed frame of the sources, where the
underlying power-law hard-\xray\ spectrum is less prone to
contamination due to any potential soft excess emission or
absorption. In each fit, the photon indices in the spectra of all
three EPIC detectors were tied to a single value, while the
normalizations were free to vary.
The best-fit $\Gamma$ values, power-law normalizations, and $\chi^2$
values from these fits are given in columns (2), (3), and (4) of
Table\,\ref{tab_properties}, respectively, and the data, their joint,
best-fit spectra, and residuals appear in Fig.\,\ref{fig_spectra}.

We also searched for intrinsic absorption in each source by jointly
fitting the spectra with a Galactic-absorbed power law model including
an intrinsic (redshifted) neutral-absorption component with solar
abundances in the same energy range quoted above.
No significant intrinsic absorption was detected in any of the
sources; upper limits on intrinsic \nh\ values appear in column (5) of
Table\,\ref{tab_properties}. Each panel of Fig.\,\ref{fig_spectra}
includes a \hbox{$\Gamma$-\nh} confidence-contour plot from this
fitting for each source. By applying $F$-tests between the models
including intrinsic absorption and those that exclude it, we found
that none of the spectra require an intrinsic absorption component.
The remarkably flat hard-\xray\ spectrum of LBQS~0109$+$0213 (with
$\Gamma=1.23$; see Table~\ref{tab_properties}) motivated an
alternative modeling, searching for an indication of partial
covering. We found that when the spectrum is fitted with a redshifted
partial-covering fraction absorber ({\sc zpcfabs} model in {\sc
  XSPEC}) and a Galactic-absorbed power law, the covering fraction is
consistent with zero and the fit is not improved with respect to a
Galactic-absorbed power law model. Therefore, the unusually flat
hard-\xray\ spectrum of LBQS~0109$+$0213 cannot be explained as being
due to partial covering.

\subsection{The \chandra\ Spectrum of HE~0926$-$0201}
\label{sec_chandra}

\begin{figure}
\centering
\plotone{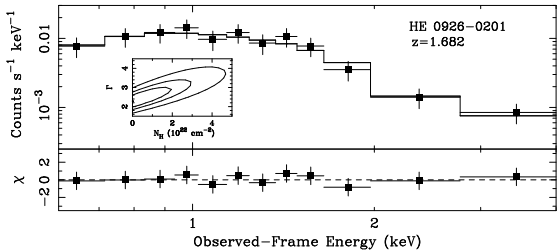}
\caption{The \chandra\ spectrum of HE~0926$-$0201. The solid line
  represents the best fit Galactic-absorbed power law model. The
  $\chi$ residuals are in units of $\sigma$ with error bars of size
  1. The inset shows 68\%, 90\%, and 99\% confidence contours for
  $\Gamma$ and \nh, when the data are fitted with an additional
  neutral intrinsic-absorption component.}
\label{fig_HE0926_chandra_spec}
\end{figure}

HE~0926$-$0201 is a highly luminous RQ AGN at $z=1.682$ for which
near-infrared spectroscopy of the \hb\ region is presented in Sulentic
\et (2006).
This source is serendipitously detected in an 8.8\,ks \chandra/ACIS-S
observation from 2002 February 20 (ID~3139) with 126 source counts in
the 0.5--8~keV observed-frame band. We analyzed the \chandra\
observation using standard {\sc ciao\footnote{\chandra\ Interactive
    Analysis of Observations. See http://asc.harvard.edu/ciao/} v3.2}
routines.
The spectrum was extracted using {\sc psextract} from a circular
aperture with a radius of eight pixels, corresponding to $\sim$4$''$,
and the events were grouped to have a minimum of 10 counts per bin
(the extremely low \chandra\ background was determined from a
source-free annular region with inner and outer radii of 10$''$ and
25$''$, respectively, centered on the source).
The spectrum was fitted over the $\gtsim2$\,keV rest-frame energy band
using {\sc xspec} with a model including a power law and a
Galactic-absorption component with a column density of \nh=3.17$\times
10^{20}$~cm$^{-2}$.
The data were also fitted with an additional intrinsic-absorption
component, similar to the procedure described above for the \xmm\
observations, and we found that the data do not warrant any additional
absorption.
The best-fit spectral parameters as well as the upper limit on the
intrinsic absorption are included in Table~\ref{tab_properties}, and
the \chandra\ spectrum and best-fit model appear in
Fig.~\ref{fig_HE0926_chandra_spec}.

\section{Results and Discussion}
\label{sec_results}

\subsection{Spectral and Temporal Properties}
\label{sec_properties}

\subsubsection{Optical Luminosities and FWHM(\hb)}
\label{sec_L5100_Hb}

Basic optical spectroscopic properties of the new high-redshift sample
are given in Table\,\ref{tab_properties}. The monochromatic luminosity
at a rest-frame wavelength of 5100\,\AA\ [$\nu L_{\nu}
(5100\,\mbox{\AA})$] is given in column (6), and FWHM(\hb) is given in
column (7); except for HE\,0926$-$0201, for which the optical data
were obtained from Sulentic \et (2006), these data were obtained from
Netzer \et (2007), who recently presented new and improved
spectroscopic measurements for all the Shemmer \et (2004) sources.
The \mbh\ and \lledd\ values in columns (8) and (9), respectively,
were determined as in S06, using the $\nu L_{\nu} (5100\,\mbox{\AA})$
and FWHM(\hb) values given in columns (6) and (7), respectively, and
based on the recent reverberation-mapping results of Peterson \et
(2004) and the Kaspi \et (2005; hereafter K05) BELR size-luminosity
relation (see also Kaspi \et 2000); the general expression we use for
\lledd\ is of the form \lledd$\propto L^{0.3} {\rm
  FWHM(H}\beta)^{-2}$. We note that the K05 relation relies on a
sample of AGNs having luminosities up to \hbox{$\nu L_{\nu}
  (5100\,\mbox{\AA})$ $\approx$10$^{46}$\,ergs\,s$^{-1}$}, and
extrapolating it to higher luminosities, such as those of ten of the
sources presented here, is somewhat uncertain; a reverberation-mapping
effort is underway to test the validity of such extrapolations (see
e.g., Kaspi \et 2007).
In addition, Bentz \et (2006, 2007) have recently suggested that
subtraction of host-galaxy starlight from the AGN optical continuum
may result in a somewhat flatter slope (e.g., $\alpha=0.54$ compared
with $\alpha=0.69$ in K05) for the BELR size-luminosity relation
across the entire K05 luminosity range. We test the possibility of
using a flatter slope below, but we note that the Bentz \et
measurements were performed for sources with $\nu L_{\nu}
(5100\,\mbox{\AA}) \ltsim$10$^{44}$\,ergs\,s$^{-1}$, while all the
sources studied here have $\nu L_{\nu} (5100\,\mbox{\AA})
\gtsim$10$^{44}$\,ergs\,s$^{-1}$ (with a median luminosity of $\nu
L_{\nu} (5100\,\mbox{\AA}) \sim$10$^{45}$\,ergs\,s$^{-1}$;
Fig.~\ref{fig_hist}). We also expect a much smaller fractional amount
of host contamination in the most luminous sources in our sample. In
light of all this, we have computed \mbh\ values using the K05 slope
across our entire luminosity range. Moreover, since in this work we
employ nonparametric statistical ranking tests, our main results are
not significantly sensitive to the precise value of the slope of the
BELR size-luminosity relation in the range $\alpha=0.54-0.69$.
To derive the normalized accretion rates, we have employed the
luminosity-dependent bolometric correction method of Marconi \et
(2004); the bolometric correction factor for the monochromatic
luminosity at 5100\,\AA\ [i.e., \lbol/$\nu L_{\nu}$(5100\,\AA)] is
\hbox{$\sim6-8$} \hbox{($\simeq5$)} for the moderate-luminosity
$z<0.5$ (high-luminosity $z=1.3-3.2$) sources in this work (see S06
for more details).

\subsubsection{Optical--X-ray Spectral Slopes}
\label{sec_aox_data}

The optical--\xray\ spectral slopes (\aox) in column (10) of
Table~\ref{tab_properties} are defined as \\ \aox$=\log(f_{\rm
  2\,keV}/f_{2500\mbox{\rm\,\scriptsize\AA}})/ \log(\nu_{\rm
  2\,keV}/\nu_{2500\mbox{\rm\,\scriptsize\AA}})$, where $f_{\rm
  2\,keV}$ and $f_{2500\mbox{\rm~\scriptsize\AA}}$ are the flux
densities at 2\,keV and 2500\,\AA, respectively. The \aox\ values were
derived using the photon indices and fluxes in columns (2) and (3),
respectively, and the optical luminosities in column\,(6), assuming a
UV continuum of the form $f_{\nu}\propto \nu^{-0.5}$ (Vanden Berk \et
2001). The \aox\ values for all of our sources are consistent with the
expected values, given their optical luminosities (e.g., Steffen \et
2006).

\subsubsection{Compton Reflection and \Ka\ Emission}
\label{sec_compton}

The relatively high redshifts of our sources allowed us to search for
a Compton-reflection continuum as well as \Ka\ emission in their \xmm\
spectra. A Compton-reflection continuum may be observed in AGNs within
the rest-frame $\sim7-60$\,keV energy range, peaking at rest-frame
$\sim$30\,keV, and it presumably originates from reflection of
hard-\xray\ photons off the relatively colder outskirts of the
accretion disk and/or the torus. We also searched for signatures of a
neutral narrow \Ka\ emission line at rest-frame 6.4\,keV, as this is
expected to appear in conjunction with a Compton-reflection
continuum. The search for continua reflected from neutral material was
carried out by fitting all our \xmm\ spectra in the \gtsim2\,keV
rest-frame energy range with {\sc xspec}, employing a
Galactic-absorbed power-law and a Compton-reflection continuum model
(i.e., the {\sc pexrav} model in {\sc xspec}; Magdziarz \& Zdziarski
1995) simultaneously with a redshifted Gaussian emission line model
(using the {\sc zgauss} model in {\sc xspec}); the Gaussian rest-frame
energy and width were fixed at $E=6.4$\,keV and $\sigma=0.1$\,keV,
respectively. We also included in the analysis the five luminous,
high-redshift sources from S06 that have \xmm\ spectra and \hb\
spectroscopy, namely PG\,1247$+$267, Q\,1346$-$036, PG\,1630$+$377,
PG\,1634$+$706, and HE\,2217$-$2818; the luminosities and accretion
rates of these sources are comparable to those of the sources
presented in this work.
Table~\ref{tab_compton} lists the best-fit parameters from these
fits. Column (3) gives the rest-frame equivalent width (EW; throughout
the paper, EWs refer to rest-frame values) of the \Ka\ emission line
and column (4) gives the relative-reflection component ($R$) of the
Compton-reflection continuum expressed as $R=\Omega/2\pi$, where
$\Omega$ is the solid angle subtended by the continuum source.

Except for the cases of PG\,1247$+$267 and PG\,1634$+$706, $F$-tests
carried out between the results of the {\sc pexrav}+{\sc zgauss} fits
and the Galactic-absorbed power-law fits (\S\,\ref{sec_xmm}) have
indicated that the new (reflection) model did not improve the fits
with respect to the corresponding Galactic-absorbed power-law fits.
Table~\ref{tab_compton} shows that, except for PG\,1630$+$377, we have
not detected any neutral \Ka\ emission in any of the sources, and
instead, we have placed upper limits on the EWs of such emission.
Although the EW(\Ka) for PG\,1630$+$377 has a nonzero lower limit, and
the EW value we found is consistent with the one found for this source
by Jim{\'e}nez-Bail{\'o}n \et (2005), we argue that the \xmm\ data for
PG\,1630$+$377 do not warrant the existence of a reflection spectrum
based on the $F$-test described above.

We have also tested whether the \xmm\ spectra of our sources exhibit
hard-\xray\ emission reflected from ionized material. This was
performed by employing the same {\sc xspec} fitting routine described
above for the case of reflection from neutral material, except that
the {\sc pexrav} model was replaced with the {\sc pexriv} model
(Magdziarz \& Zdziarski 1995), and the energy of the \Ka\ line was
free to vary in the 6.7--6.97 rest-frame energy range to account for
an ionized iron line. The results of these fits, i.e., $R$ parameters
and EW(\Ka) values, are consistent with those presented in
Table~\ref{tab_compton} for the case of reflection from neutral
material (even though the energies of the \Ka\ lines were required to
be somewhat higher).

The non-detections of \Ka\ emission lines in the highly luminous
sources under study are in accord with previous studies claiming that
the strength of such emission lines decreases with increasing
luminosity (aka the `\xray\ Baldwin Effect'; e.g., Iwasawa \&
Taniguchi 1993; Page \et 2004a; Zhou \& Wang 2005; Bianchi \et
2007). However, the relatively high upper limits we obtain for our
EW(\Ka) values cannot rule out the possibility that such an `\xray\
Baldwin Effect' is very weak or does not even exist (e.g.,
Jim{\'e}nez-Bail{\'o}n \et 2005; Jiang \et 2006). Given the high
luminosities of our sources, the expected EW(\Ka) values from such an
EW-$L$ relation are of order $\sim$50\,eV, while in most cases, our
upper limits are considerably higher (see Table~\ref{tab_compton} and
e.g., Bianchi \et 2007).

We detect statistically significant Compton-reflection continua only
in PG\,1247$+$267 and PG\,1634$+$706 (Table~\ref{tab_compton}). The
$R$ value we find for the first of these sources is consistent with
the one found by Page \et (2004b) for the same \xmm\
spectrum. However, in contrast with our results, Page \et (2004b) have
not detected a significant Compton-reflection feature in the same
\xmm\ spectrum of PG\,1634$+$706 that we have analyzed, while the
upper limit we obtain on the EW(\Ka) for the source is consistent with
their finding; we also note that Nandra \et (1995) have not detected
any reflection features in an {\sl ASCA} spectrum of the source. In
addition, Page \et (2004b) {\em have} detected an \Ka\ line in
PG\,1247$+$267 while we have not. Interestingly, inspection of Fig.~1
of Page \et (2004b) suggests the existence of a reflection component
in both sources. An alternative spectral fitting, including a thermal
component and a power-law spectrum for each source, was recently
employed by Ruiz \et (2007); their study finds a soft excess and no
reflection component in both sources.
These partial discrepancies may be a consequence of the different
modelings and different energy ranges used in the different
studies. The apparently non-physical $R$ values we obtain for
PG\,1247$+$267 and PG\,1634$+$706 (i.e., $\Omega>4\pi$ steradians) may
be due to the effects of general relativistic light bending (e.g.,
Fabian \et 2002; Fabian \& Vaughan 2003).

To improve our ability to detect or constrain the strengths of any
reflection features, we jointly fitted the \xmm\ spectra of our nine
high-redshift sources from Table~\ref{tab_compton}. In the joint
fitting, we used the {\sc pexrav}+{\sc zgauss} models as before and
tied all the $R$ values and normalizations to a single value (the flux
level of each source was controlled by assigning to it a scaling
factor that was allowed to vary freely, and the photon index for each
source was also free to vary). We ran the joint fitting process four
times. In the first run, all nine sources were considered. In the
second run, we removed the two sources from Table~\ref{tab_compton} in
which a significant Compton-reflection continuum was detected. In the
third run, we removed all sources in which a non-zero $R$ value was
detected, and in the fourth run we removed PG\,1630$+$377 in which an
\Ka\ line was detected.

The best-fit values of EW(\Ka) and $R$ from the joint-fitting process
appear in Table~\ref{tab_compton_joint_fit} where we also provide the
total number of counts used in the fit and the mean redshift of the
sources considered in each run. By comparing the results in
Table~\ref{tab_compton_joint_fit} with those in
Table~\ref{tab_compton}, one can see that when all sources are
considered (run\,I), a large value of $R$ is detected, while a tight
constraint is placed on the mean EW(\Ka); this result is dominated by
the properties of PG\,1247$+$267 and PG\,1634$+$706 that comprise
$\sim$60\% of the counts. In run\,II, one can see that despite the
removal of all sources in which a Compton-reflection continuum was
detected, a mean reflection component is detected, although with a
relatively small value of $R$. In run\,III, the \Ka\ detection is
probably due to the fact that PG\,1630$+$377 comprises $\sim$30\% of
the total counts; no significant Compton reflection is detected in
this case. In the final run, no significant mean reflection emission
is detected; the constraint on $R$ is relatively tight while the
constraint on EW(\Ka) is rather weak.

\begin{deluxetable*}{lccccccccc}
\tablecolumns{10}
\tabletypesize{\scriptsize}
\tablewidth{0pc}
\tablecaption{Best-Fit X-Ray Spectral Parameters and Optical
  Properties \label{tab_properties}}
\tablehead{
\colhead{} &
\colhead{} &
\colhead{} &
\colhead{} &
\colhead{} &
\colhead{$\log \nu L_{\nu}(5100\,\mbox{\AA})$} &
\colhead{FWHM(\hb)} &
\colhead{$\log M_{\rm BH}$} &
\colhead{} &
\colhead{} \\
\colhead{{\sc AGN}} &
\colhead{{\sc $\Gamma$}} &
\colhead{$f_{\nu}$(1\,keV)\tablenotemark{a}} &
\colhead{$\chi^{2}/\nu$} &
\colhead{\nh\tablenotemark{b}} &
\colhead{(ergs\,s$^{-1}$)} &
\colhead{(\kms)} &
\colhead{(\msun)} &
\colhead{\lledd} &
\colhead{\aox} \\
\colhead{(1)} &
\colhead{(2)} &
\colhead{(3)} &
\colhead{(4)} &
\colhead{(5)} &
\colhead{(6)} &
\colhead{(7)} &
\colhead{(8)} &
\colhead{(9)} &
\colhead{(10)}
}
\startdata
LBQS\,0109$+$0213 & $1.23^{+0.12}_{-0.12}$ & $6.1^{+0.8}_{-0.7}$ &
70/53 & $\le1.46$ & 46.8 & 7959 & 10.4 & 0.1 &
$-$1.89 \\
2QZ\,J023805.8$-$274337 & $2.13^{+0.16}_{-0.15}$ & $7.8^{+0.8}_{-0.8}$
& 47/39 & $\le3.71$ & 46.6 & 3403 & 9.5 & 0.5 &
$-$1.66 \\
Q\,1318$-$113 & $1.96^{+0.07}_{-0.07}$ & $24.8^{+1.2}_{-1.2}$ & 70/62
& $\le0.38$ & 46.9 & 4665
& 10.0 & 0.3 & $-$1.64 \\
SBS\,1425$+$606 & $1.76^{+0.14}_{-0.13}$ & $13.6^{+1.7}_{-1.7}$ &
38/39 & $\le2.33$ & 47.4 & 4964 & 10.4 & 0.4
& $-$1.82 \\
HE\,0926$-$0201 & $2.28^{+0.36}_{-0.34}$ & $21.8^{+3.8}_{-3.8}$ & 2/10
& $\le1.72$ & 47.0\tablenotemark{c} & 5100\tablenotemark{c} & 10.1 &
0.3 & $-$1.80
\enddata
\tablecomments{The best-fit photon index, normalization, and $\chi^2$
  were obtained from a Galactic-absorbed power-law model. Errors
  represent 90\% confidence limits, taking one parameter of interest
  ($\Delta \chi^{2}=2.71$). The optical data in columns 6 and 7 were
  obtained from Shemmer \et (2004) and Netzer \et (2007) except for
  HE\,0926$-$0201.}
\tablenotetext{a}{Power-law normalization given as the flux density at
  an observed-frame energy of 1\,keV with units of
  10$^{-32}$\,ergs\,cm$^{-2}$\,s$^{-1}$\,Hz$^{-1}$; except for
  HE\,0926$-$0201, this refers to the pn data, taken from joint
  fitting of all three EPIC detectors with the Galactic-absorbed
  power-law model.}
\tablenotetext{b}{Intrinsic column density in units of
  10$^{22}$\,cm$^{-2}$. Upper limits were computed with the
  intrinsically absorbed power-law model with Galactic absorption, and
  represent 90\% confidence limits for each value.}
\tablenotetext{c}{Obtained from Sulentic \et (2006).}
\end{deluxetable*}

These results do not provide a clear picture for the dependence of the
reflection spectrum on luminosity or the accretion rate. Most of our
high-redshift sources do not exhibit significant reflection
components, as might be expected given their very high
luminosities. Only two of these sources show a significant Compton
`hump', and a strong \Ka\ line is detected in another. In addition,
significant detections of Compton humps are not accompanied by
corresponding detections of \Ka\ lines.
The results of the joint-fitting process also portray a mixed picture
for the average reflection spectrum of luminous, high-accretion rate
RQ AGNs as a class. The relatively weak constraints on the mean
reflection properties of this class do not allow us to either confirm
or rule out the existence of an `\xray\ Baldwin Effect', regardless of
whether the \xray\ luminosity or the accretion rate drives the
anticorrelation with EW(\Ka) (e.g., Zhou \& Wang 2005; Bianchi \et
2007).

\subsubsection{Soft Excesses}
\label{sec_soft_excess}

By extending the \xray\ spectral fitting to the entire EPIC energy
range (0.2--10\,keV), we checked whether any of our nine high-redshift
sources with \xmm\ spectra (including the five high-redshift sources
from S06) shows evidence for excess soft-\xray\ emission, frequently
observed in lower-redshift AGNs (e.g., Comastri \et 1992; Reeves \&
Turner 2000; Piconcelli \et 2005). While the physical nature of the
soft excess is uncertain its presence is more pronounced among local
high-accretion rate AGNs, i.e., narrow-line Seyfert~1 (NLS1) galaxies
(e.g., Vaughan \et 1999a; Boller \et 2002; Czerny \et 2003; Vignali
\et 2004); hence it is of interest to search for its existence in our
luminous high-accretion rate sources at high redshift. We extrapolated
the best-fit Galactic-absorbed power-law model obtained for rest-frame
energies \gtsim2\,keV (see \S\,\ref{sec_xmm} and
Table~\ref{tab_properties}) to the 0.2--10~keV observed-frame energy
range and no signature of soft excess emission was detected in any of
the sources (i.e., no systematic residuals were present for the
extrapolated fits).

This result is not unexpected given the relatively high redshifts of
our sources and the low-energy cutoff (0.2\,keV) of the EPIC cameras.
For example, Porquet \et (2004) have found that the effective
temperatures of the soft excess-components in a sample of
moderate-luminosity sources at $z<0.5$ are of the order of
$\sim0.2-0.3$~keV. For our ten sources with $z\sim2$, such
temperatures shift to observed-frame values of
$\sim0.1$~keV. Furthermore, since the soft-excess component in most
type~1 AGNs is typically observed below a rest-frame energy of
$\sim0.7$~keV (except for some high \lledd\ NLS1s, in which this
component may extend up to 1.5~keV; e.g., Vaughan \et 1999b), the
detection of such a component with \xmm\ is challenging even for
sources with $z\sim1$. Nevertheless, we have placed upper limits on
the strength of a potential soft-excess component in two of our ten
luminous sources that have the lowest redshifts, namely PG\,1630$+$377
and PG\,1634$+$706 (with $z=1.476$ and $z=1.334$, respectively). In
these two cases, the extension of the energy range to 0.2\,keV has
increased the number of photons in the fit by $\sim50$\%. The
constraint on the data-to-model ratio in the \ltsim2\,keV rest-frame
band, as a result of the extrapolation of the best-fit slope to the
entire EPIC energy range, is \ltsim2.0 and \ltsim1.5, respectively.
Finally, we note that the non-detection of soft-excess emission in
PG\,1247$+$267, PG\,1630$+$377, and PG\,1634$+$706 is consistent with
previous studies of their \xmm\ spectra (Page \et 2004b; Piconcelli
\et 2005).

\subsubsection{X-ray Variability}
\label{sec_variability}

Two of our sources, SBS\,1425$+$606 and LBQS\,0109$+$0213, were
previously detected by the {\sl ROSAT} PSPC and HRI, respectively. For
the first of these, we measured an unabsorbed flux of
$4.7\times10^{-14}$\,ergs\,cm$^{-2}$\,s$^{-1}$ in the 0.5--2.0\,keV
observed-frame band from our \xmm\ data, which is somewhat higher
than, but consistent (within the errors) with the {\sl ROSAT} flux
(see also Just \et 2007). Similarly, we measured for the second source
an unabsorbed flux of $2.0\times10^{-14}$\,ergs\,cm$^{-2}$\,s$^{-1}$
in the 0.5--2.0\,keV observed-frame band, which is a factor of $\sim3$
lower than the value calculated from the {\sl ROSAT} count rate. In
both cases, we assumed that the photon index we measured from the
\xmm\ observations at rest-frame energies of $\gtsim 2$\,keV also
extends to the {\sl ROSAT} bandpass (i.e., down to energies of
0.1\,keV in the observed frame), based on the lack of detectable soft
excesses in these sources (\S~\ref{sec_soft_excess}).
Although highly luminous RQ AGNs such as those presented here are not
expected to exhibit pronounced \xray\ variations even on timescales of
several years (e.g., Lawrence \& Papadakis 1993), the case of
LBQS\,0109$+$0213 is not unique, since several other luminous sources
at high redshifts have displayed \xray\ variations with similar
amplitudes (e.g., Paolillo \et 2004; Shemmer \et 2005).

\begin{deluxetable}{lccc}
\tablecolumns{4}
\tablewidth{0pc}
\tablecaption{Compton Reflection and Iron
  Emission \label{tab_compton}}
\tablehead
{
\colhead{} &
\colhead{} &
\colhead{EW(\Ka)\tablenotemark{a}} &
\colhead{} \\
\colhead{AGN} &
\colhead{$z$} &
\colhead{(eV)} &
\colhead{$R$\tablenotemark{b}}
}
\startdata
LBQS\,0109$+$0213 & 2.349 & $\leq252$ & $3.3^{+39.0}_{-3.0}$ \\
2QZ\,J023805.8$-$274337 & 2.471 & $\leq375$ & $\leq1.4$ \\
PG\,1247$+$267\tablenotemark{c} & 2.038 & $\leq188$ & $2.9^{+38.8}_{-2.0}$ \\
Q\,1318$-$113 & 2.306 & $\leq52$ & $1.4^{+3.0}_{-1.2}$ \\
Q\,1346$-$036 & 2.370 & $\leq298$ & $\leq7.2$ \\
SBS\,1425$+$606 & 3.202 & $\leq613$ & $\leq1.9$ \\
PG\,1630$+$377 & 1.476 & $458^{+384}_{-374}$ & $\leq6.7$ \\
PG\,1634$+$706\tablenotemark{c} & 1.334 & $\leq49$ & $3.0^{+3.2}_{-1.3}$ \\
HE\,2217$-$2818 & 2.414 & $\leq233$ & $\leq1.6$
\enddata
\tablecomments{Best-fit parameters of fitting each spectrum at the
  \gtsim2\,keV rest-frame energy range with a model consisting of a
  Galactic-absorbed power-law, a Compton-reflection component, and a
  neutral \Ka\ emission line. Errors represent 90\% confidence limits,
  taking one parameter of interest ($\Delta \chi^{2}=2.71$).}
\tablenotetext{a}{Rest-frame equivalent width of a neutral \Ka\
  emission line at rest-frame $E=6.4$\,keV and a fixed width of
  $\sigma=0.1$\,keV.}
\tablenotetext{b}{Relative Compton-reflection parameter; see text for
  details.}
\tablenotetext{c}{$F$-test indicates that the {\sc pexrav}+{\sc
    zgauss} model results in an improved fit with respect to a
  power-law model.}
\end{deluxetable}

We also searched for rapid (on timescales of $\sim$1\,hr in the rest
frame) \xray\ variations in the \xmm\ and \chandra\ data of our five
new sources by applying Kolmogorov-Smirnov tests to the lists of
photon arrival times from the event files, but no significant
variations were detected. Together with the results of S06, this
suggests that high accretion rates do not necessarily lead to faster
and higher-amplitude \xray\ flux variations, as has been previously
expected based on \xray\ variability properties of some NLS1s, hence
spoiling the analogy between NLS1s and luminous, high-redshift AGNs
(e.g., Grupe \et 2006). On the other hand, our results are consistent
with the idea that more massive BHs lead to longer timescales and
smaller amplitudes of \xray\ flux variations (e.g., O'Neill \et 2005).

\subsection{The Hard-X-ray Power-Law Photon Index}
\label{sec_hardX}

\subsubsection{Breaking the FWHM(\hb)-\lledd\ Degeneracy}
\label{sec_degeneracy}

\begin{figure*}
\epsscale{1.0}
\centering
\plotone{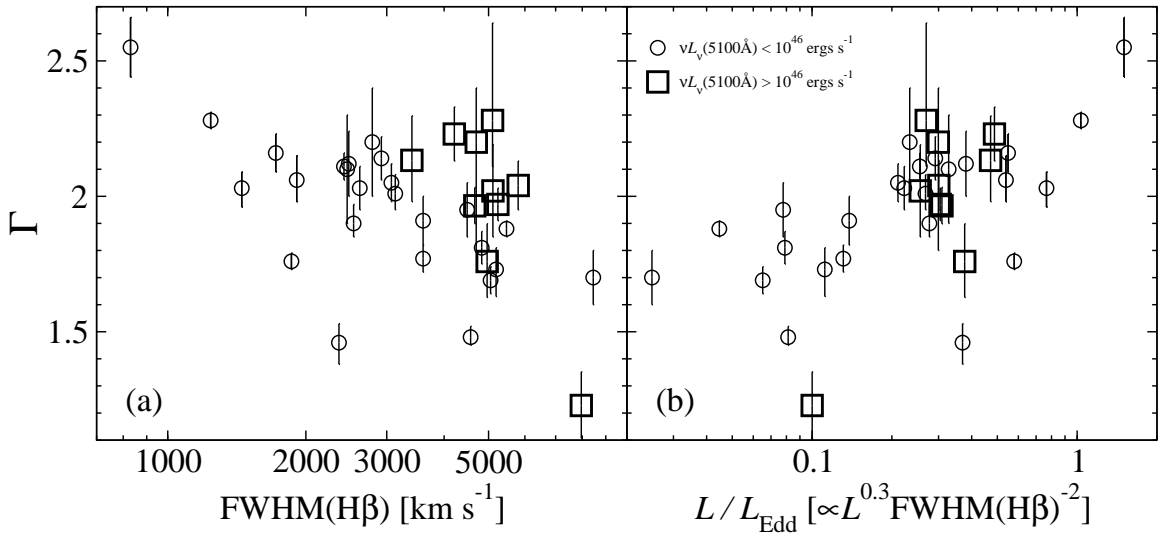}
\caption{The hard-\xray\ photon index vs. FWHM(\hb) ({\it a}) and
  \lledd\ ({\it b}). Circles mark sources at $z<0.5$ with $\nu
  L_{\nu}$(5100\AA)$<10^{46}$\,ergs\,s$^{-1}$ from the Piconcelli \et
  (2005) sample. Squares mark luminous sources at $z=1.3-3.2$ with
  $\nu L_{\nu}$(5100\AA)$>10^{46}$\,ergs\,s$^{-1}$. Error bars on
  $\Gamma$ are shown at the 90\% confidence level.}
\label{fig_GE}
\end{figure*}

In Fig.\,\ref{fig_GE} we plot $\Gamma$ versus FWHM(\hb) and \lledd\
for our core sample of 35 AGNs. Twenty five of these sources are
Palomar Green (PG) quasars (Schmidt \& Green 1983) at $z<0.5$ with
$44\ltsim \log [\nu L_{\nu} (5100~\mbox{\AA})] \ltsim 46$, and ten are
at $z=1.3-3.2$ with $46\ltsim \log [\nu L_{\nu} (5100~\mbox{\AA})]
\ltsim 48$.
In S06, it was shown that both FWHM(\hb) and \lledd\ are significantly
correlated with $\Gamma$ when only the 25 moderate-luminosity
($z<0.5$) sources are considered (in agreement with Porquet \et 2004;
Piconcelli \et 2005; see also Table~\ref{tab_corr_coeff}).
This FWHM(\hb)-\lledd\ degeneracy emerges as a consequence of
considering only sources with a relatively narrow luminosity range,
i.e., $44\ltsim \log [\nu L_{\nu} (5100~\mbox{\AA})] \ltsim 46$ (see
also \S\,\ref{sec_L5100_Hb} and S06).
When the ten highly luminous AGNs at \hbox{$z=1.3-3.2$} are introduced
to the analysis, both FWHM(\hb) and \lledd\ remain significantly
correlated with $\Gamma$ (with $>99$\% confidence); however, the
significance (in terms of chance probability) of the
$\Gamma$-FWHM(\hb) correlation drops considerably while the
significance of the $\Gamma$-\lledd\ correlation increases
(Table~\ref{tab_corr_coeff}). The results of the $\Gamma$-\lledd\
correlation do not significantly change when a constant bolometric
correction factor of 7 (which is the average correction factor for the
luminosity range of our sample; see \S\,\ref{sec_L5100_Hb}) is used to
derive \lledd\ values; in this case, the significance of the
correlation increases with the chance probability dropping from
$p=1.6\times 10^{-3}$ to $p=4\times 10^{-4}$.
As suggested in S06, the extension of the luminosity range of the
sample to high luminosities has allowed us to break the degeneracy
between the $\Gamma$-FWHM(\hb) and $\Gamma$-\lledd\ correlations and
show that \lledd\ drives the correlation with $\Gamma$.

\begin{deluxetable}{lccccc}
\tablecolumns{6}
\tablewidth{0pc}
\tablecaption{Compton Reflection and Iron
  Emission - Joint Fitting \label{tab_compton_joint_fit}}
\tablehead
{
\colhead{Run} &
\colhead{No. of} &
\colhead{No. of} &
\colhead{} &
\colhead{EW(\Ka)\tablenotemark{a}} &
\colhead{} \\
\colhead{No.} &
\colhead{Sources} &
\colhead{photons} &
\colhead{$\left < z \right >$} &
\colhead{(eV)} &
\colhead{$R$}
}
\startdata
I\tablenotemark{b} & 9 & 39,173 & 2.22 & $\leq27$ & $2.2^{+1.1}_{-0.8}$ \\
II\tablenotemark{c} & 7 & 15,307 & 2.37 & $\leq114$ & $0.9^{+0.9}_{-0.6}$ \\
III\tablenotemark{d} & 5 & 9,814 & 2.39 & $150^{+133}_{-109}$ & $\leq1.4$ \\
IV\tablenotemark{e} & 4 & 7,112 & 2.61 & $\leq300$ & $\leq1.1$
\enddata
\tablecomments{Best-fit parameters of joint fitting the spectra at the
  \gtsim2\,keV rest-frame energy range with a model consisting of a
  Galactic-absorbed power-law, a Compton-reflection component, and a
  neutral \Ka\ emission line. Errors represent 90\% confidence limits,
  taking one parameter of interest ($\Delta \chi^{2}=2.71$).}
\tablenotetext{a}{The EW(\Ka) corresponds to a rest-frame at the
  given $\left < z \right >$.}
\tablenotetext{b}{All sources from Table~\ref{tab_compton}.}
\tablenotetext{c}{All sources from Table~\ref{tab_compton}, excluding
  PG\,1247$+$267 and PG\,1634$+$706.}
\tablenotetext{d}{All sources from Table~\ref{tab_compton}, excluding
  PG\,1247$+$267, PG\,1634$+$706, LBQS\,0109$+$0213, and
  Q\,1318$-$118.}
\tablenotetext{e}{All sources from Table~\ref{tab_compton}, excluding
  PG\,1247$+$267, PG\,1630$+$377, PG\,1634$+$706, LBQS\,0109$+$0213,
  and Q\,1318$-$118.}
\end{deluxetable}

We have also repeated the Mann-Whitney (MW) nonparametric rank test
performed in S06 on all sources with FWHM(\hb) values that lie in the
range 3400$<$FWHM(\hb)$<$8000\,\kms\ [which is the FWHM(\hb) interval
of the ten luminous sources in our sample]. We find a significant
deviation (with $>99.5$\% confidence) between the $\Gamma$ values of
the ten luminous sources and the nine moderate-luminosity
\hbox{($z<0.5$)} sources in this range.
In contrast, when the MW test is performed in the
0.1\ltsim\lledd\ltsim0.5 range, the $\Gamma$ values of the two groups
of AGNs are not significantly different (as can be seen clearly from
Fig.\,\ref{fig_GE}b). These results reinforce the S06 argument that
$\Gamma$ depends primarily on \lledd.

We note that in the case of two of our luminous sources,
LBQS~0109$+$0213 and SBS~1425$+$606, the $\Gamma$ values are
consistent with the expected values of both FWHM(\hb) and \lledd, and
hence they do not assist in removing the FWHM(\hb)-\lledd\
degeneracy. The first of these sources has a remarkably flat
hard-\xray\ spectral slope (with $\Gamma=1.23$; see \S~\ref{sec_xmm}
and Table~\ref{tab_properties}), which was not expected given its
original \lledd\ determination from Shemmer \et (2004; see their
Table~2).
Owing to the new and improved spectroscopic measurements of Netzer \et
(2007) for LBQS~0109$+$0213 (SBS~1425$+$606), FWHM(\hb) has increased
from 5781\,\kms\ to 7959\,\kms\ (3144\,\kms\ to 4964\,\kms), and hence
\lledd\ has decreased from 0.2 to 0.1 (0.9 to 0.4). For the other
Shemmer \et (2004) sources used in this work, the Netzer \et (2007)
FWHM(\hb) measurements are consistent with those of Shemmer \et
(2004). The optical data presented in Table~\ref{tab_properties} are
based on the new Netzer \et (2007) measurements.

Our result that $\Gamma$ depends more strongly on \lledd, rather than
on FWHM(\hb), may also serve as a partial vindication for the use of
the reverberation-mapping based method to determine \lledd\ (see
\S\,\ref{sec_L5100_Hb}). From a pure mathematical point of view, in
the tests described above, we essentially multiplied one observational
parameter, FWHM(\hb), with another, $\nu L_{\nu} (5100~\mbox{\AA})$,
and yet the dependence of $\Gamma$ on the product of the two
parameters (i.e., \lledd) turned out to be stronger than the
dependence on the single parameter. If the product of the two
parameters had no physical significance, then there would be no
apparent reason to expect this result.

\subsubsection{The $\Gamma$-\lledd\ Correlation}
\label{sec_GE_corr}

We use the significant correlation we find between $\Gamma$ and
\lledd\ to derive a linear relationship between the two parameters
employing a variety of statistical methods. Hereafter, we take the
error bars on $\Gamma$ at the 1\,$\sigma$ level. A standard $\chi^2$
minimization method weighted by the errors on $\Gamma$ yields the
following relation (with 1~$\sigma$ errors):
\begin{equation}
\label{eq_Gamma_lledd}
\Gamma = (0.31\pm0.01) \log \left ( L/L_{\rm Edd} \right ) + (2.11\pm0.01).
\end{equation}
The best-fit coefficients of this relation are consistent with those
obtained by Wang \et (2004) and Kelly (2007) who find a similar
correlation in low--moderate luminosity sources. However, the
$\chi^{2}$ value obtained by this minimization
\hbox{($\chi^{2}/\nu=980/33$)} suggests that either the data are not
well represented by a linear model, the error bars on $\Gamma$ are
underestimated, or there is additional intrinsic scatter in the data.
Following the methods outlined in Tremaine \et (2002) and K05, and by
assuming a $\chi^{2}/\nu=33/33$ (i.e., a reduced $\chi^{2}=1$), we
obtain an estimate of the additional potential scatter of $\Delta
\Gamma \sim0.1 \times \Gamma$ in the dependent parameter; this scatter
is larger than the typical measurement errors on $\Gamma$.

\begin{deluxetable}{lccc}
\tablecolumns{4}
\tablewidth{0pc}
\tablecaption{Correlation Coefficients and
  Significance \label{tab_corr_coeff}}
\tablehead
{
\colhead{Correlation} &
\colhead{$r_{\rm S}$} &
\colhead{$p$} &
\colhead{$N$}
}
\startdata
$\Gamma$-FWHM(\hb) & $-0.61$ & $1.2 \times 10^{-3}$ & 25 \\
$\Gamma$-FWHM(\hb) & $-0.44$ & $8.4 \times 10^{-3}$ & 35 \\
$\Gamma$-\lledd\   & \phm{--}$0.60$ & $1.5 \times 10^{-3}$ & 25 \\
$\Gamma$-\lledd\   & \phm{--}$0.55$ & $6.0 \times 10^{-4}$ & 35
\enddata
\tablecomments{The last three columns represent the Spearman-rank
  correlation coefficient, chance probability, and number of sources
  for each correlation, respectively.}
\end{deluxetable}

The observed intrinsic scatter may be induced by the uncertainty in
the BH-mass estimate, and/or additional unknown physical properties,
such as the optical depth in the corona, orientation, and BH spin,
that vary from source to source. An additional potential source for
this scatter may be attributed to variability; for example,
variability may seem to have a significant contribution to the scatter
since the \xray\ data and the rest-frame optical data (used for
obtaining \lledd) are not contemporaneous.
However, as discussed in S06, variability (i.e., changes in both
$\Gamma$ and \lledd) is not expected to dominate the observed scatter
in this correlation (see also K05; Steffen \et 2006; Wilhite \et
2007); its effects are expected to be even less significant in this
case, since our sources are mostly luminous and thus have smaller
amplitudes of \xray\ and optical flux variations (e.g., Lawrence \&
Papadakis 1993; Kaspi \et 2007).\footnote{We note that some
  low-luminosity AGNs are known to exhibit large \xray\ flux
  variations (with amplitudes of factors of $\sim$10 or more) and
  corresponding \xray\ spectral slope variations (with $\Delta \Gamma
  \sim 0.4$) that are often due to changing spectral contributions
  from reflection and/or absorption (e.g., Taylor \et 2003; Krongold
  \et 2007; Grupe \et 2008). Such changes have not been observed for
  moderate--high luminosity sources such as those under study in this
  work.}

\begin{figure*}
\centering
\plotone{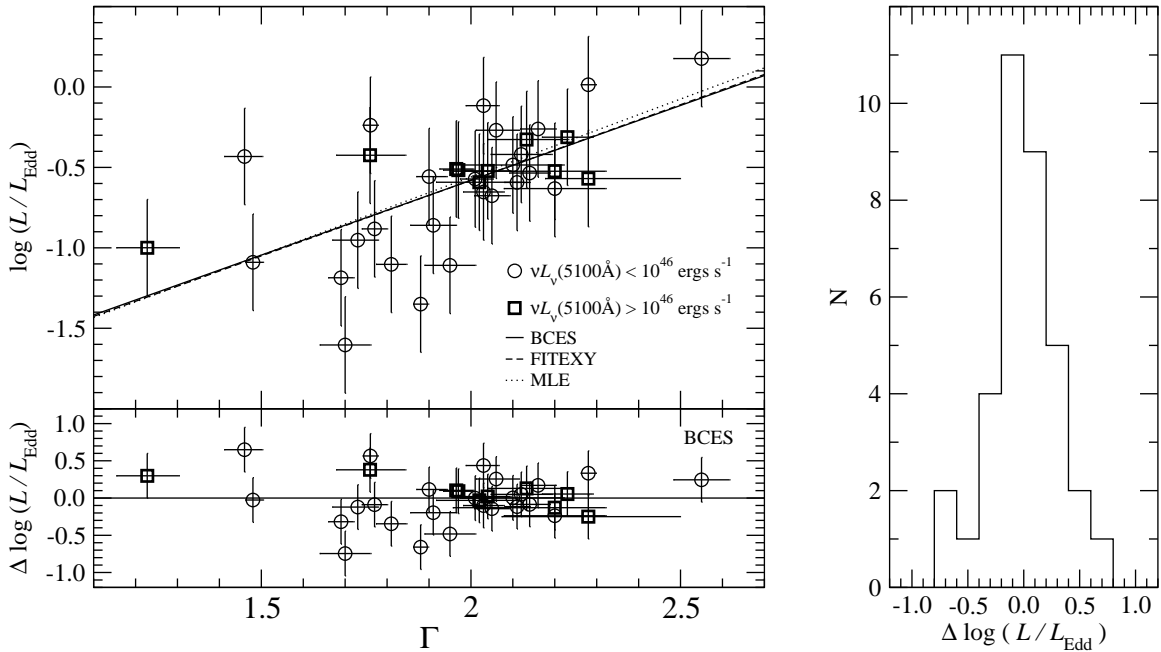}
\caption{{\it Left:} the \lledd-$\Gamma$ correlation ({\it top}) and
  residuals from the BCES fit ({\it bottom}). Symbols are identical to
  those in Fig.\,\ref{fig_GE}. Errors bars are at the $1\sigma$ level.
  The solid line, dashed line, and dotted line mark the best-fit
  linear relation from the BCES, FITEXY, and MLE methods,
  respectively, outlined in the text (note that the three lines are
  almost overlapping; see Table~\ref{tab_linear_fits}). {\it Right:}
  distribution of the $\log$(\lledd) residuals from the BCES fit for
  the entire sample.}
\label{fig_GE_yx}
\end{figure*}

To test whether $\Gamma$ can serve as an accretion-rate indicator we
switched the roles of $\Gamma$ and \lledd, allowing the first to serve
as the independent variable. To account for the scatter in the
\hbox{$\Gamma$-\lledd} correlation, we performed a linear-regression
analysis using the Bivariate Correlated Errors and Scatter method
(BCES; Akritas \& Bershady 1996) on the data. For this purpose, we
assumed that typical uncertainties on the determination of \mbh\ are a
factor of $\sim~2$ (e.g., Kaspi \et 2000), and therefore assigned
homoscedastic (i.e., uniform variance), $1\sigma$ errors of 0.3~dex on
$\log$(\lledd). The best-fit linear relation using the BCES bisector
result (with 1~$\sigma$ errors) is
\begin{equation}
\label{eq_lledd_Gamma}
\log (L/L_{\rm Edd}) = (0.9\pm0.3) \Gamma - (2.4\pm0.6).
\end{equation}
A linear regression based on $\chi^2$ minimization using the FITEXY
routine (Press \et 1992) gives consistent results to those obtained
with BCES with $\chi^{2}/\nu=34.24/33$. In addition, we also performed
a linear-regression analysis on the data using the maximum-likelihood
estimate (MLE) of Kelly (2007). The results of the different linear
fits are presented in Table~\ref{tab_linear_fits}, and are shown in
Fig.~\ref{fig_GE_yx}. In all three linear-regression methods outlined
above, the average scatter on the predicted value of $\log$(\lledd) is
$\sim0.35$ dex.
This is only slightly higher than the typical uncertainty, $\sim0.3$
dex, associated with $\log$(\lledd) determinations using the
reverberation-mapping based \mbh\ determinations.

We have also computed \lledd\ values using the Bentz \et (2007) slope
for the BELR size-luminosity relation (see
\S\,\ref{sec_L5100_Hb}). This flatter slope ($\alpha=0.54$) results in
\mbh\ values that are smaller by a factor of $\sim0.7$, on average,
and hence \lledd\ values that are larger, on average, by the same
factor. This has no noticeable effect on the strength of the
\hbox{$\Gamma$-\lledd} correlation (as expected for a nonparametric
correlation) or on its slope; only the intercept (i.e., $\alpha$
values in Table~\ref{tab_linear_fits}) is affected, and it increases
slightly within the current uncertainty on its value. We conclude that
the $\Gamma$-\lledd\ relation is not significantly affected by small
deviations from the K05 slope.

\begin{deluxetable*}{cccccc}
\tablecolumns{6}
\tablewidth{0pc}
\tablecaption{Linear Regression Coefficients for the \lledd-$\Gamma$
  correlation \label{tab_linear_fits}}
\tablehead
{
\multicolumn{2}{c}{BCES bisector} &
\multicolumn{2}{c}{FITEXY} &
\multicolumn{2}{c}{MLE \tablenotemark{a}} \\
\colhead{$\beta$} &
\colhead{$\alpha$} &
\colhead{$\beta$} &
\colhead{$\alpha$} &
\colhead{$\beta$} &
\colhead{$\alpha$}
}
\startdata
$0.93\pm0.31$ & $-2.44\pm0.63$ & $0.94\pm0.21$ & $-2.46\pm0.42$ &
$0.97\pm0.26$ & $-2.50\pm0.50$
\enddata
\tablecomments{The slope and intercept for each method are represented
  by $\beta$ and $\alpha$, respectively. Errors are at 1~$\sigma$
  confidence levels.}
\tablenotetext{a}{See Kelly (2007) for more details.}
\end{deluxetable*}

Based on our high-quality sample of 35 sources, we conclude that the
hard-\xray\ power-law spectral slope can predict the value of the
normalized accretion rate in RQ AGNs, across four orders of magnitude
in AGN luminosity, with an acceptable uncertainty level of a factor of
\ltsim3. This may offer a useful new tool to probe the history of BH
growth, based almost exclusively on the availability of high-quality
hard-\xray\ spectra of RQ AGNs; we discuss this in more detail in
\S\,\ref{sec_BHgrowth} below. Although not unexpected, and previously
predicted by several studies (see \S\,\ref{sec_introduction}), our
result is by far the most reliable indication that the shape of the
hard-\xray\ power-law spectrum is largely controlled by \lledd. As
mentioned in \S\,\ref{sec_introduction}, a possible explanation is
that the corona acts as a `thermostat' by cooling more efficiently
when the disk emission increases, manifested by a steepening of the
hard-\xray\ spectrum.

In this context, it is also interesting to note that our result may
readily explain the narrow ranges observed for the values of both
$\Gamma$ and \lledd\ in optically selected moderate--high luminosity,
type~1 RQ AGNs. While \lledd\ is relatively narrowly distributed
around a value of $\sim$0.3 (0.28 in our sample of 35 sources), with a
typical dispersion of a factor of $\sim$5 around that value (e.g.,
McLure \& Dunlop 2004; Kollmeier \et 2006; Netzer \et 2007; Shen \et
2008), $\Gamma$ values for such sources typically lie in the narrow
range of \hbox{$\sim$1.5--2.5} (1.2--2.6 for our sample of 35 sources;
see also Vignali \et 2005).
These ranges in \lledd\ and $\Gamma$ may be different for lower
luminosity sources. In particular, for sources with $\nu
L_{\nu}$(5100\AA)$\ltsim 10^{42}$\,ergs\,s$^{-1}$, a possible
difference may, in part, be attributed to a difference in their
accretion mode as compared with the moderate--high luminosity sources
studied in this work.
Finally, we point out that a possible dependence of the bolometric
correction for $L_{\rm 2-10\,keV}$ on \lledd, recently reported by
Vasudevan \& Fabian (2007), may, in part, be reflected by our results.

\subsection{What Determines the Optical--X-ray SED?}
\label{sec_SED}

The \aox\ parameter is known to have a strong anticorrelation with
optical--UV luminosity at 2500\,\AA\ [hereafter $L_{\nu}
(2500\,\mbox{\AA})$], and it shows no significant dependence on
redshift (e.g., Vignali \et 2003; Strateva \et 2005; Steffen \et 2006;
Just \et 2007; but see also Kelly \et 2007).
This is equivalent to a non-linear relation between \xray\ and
optical--UV luminosity of the form \hbox{$L_{\nu} (2\,{\rm keV})
  \propto L_{\nu} (2500\,\mbox{\AA})^{\alpha}$}, where $\alpha<1$ (see
Eq.\,7 of Just \et 2007 who find \hbox{$\alpha=0.709\pm0.010$}).
This relation still lacks a sound physical interpretation and, in
particular, it is not clear what mechanism controls the proportion of
reprocessed hard-\xray\ emission from the corona with respect to the
UV emission from the disk.

In order to test whether \aox\ depends on \lledd, we selected 81
sources from the Steffen \et (2006) sample with available FWHM(\hb)
measurements (all these have $z\ltsim0.75$ and they include the 25
sources with $z<0.5$ from our core sample; \S\,\ref{sec_degeneracy}),
and complemented this sample with the ten luminous sources at
$z=1.3-3.2$ from S06 and this work. For about half of the Steffen \et
(2006) sub-sample, FWHM(\hb) values were obtained from Boroson \&
Green (1992), and for the rest we used FWHM(\hb) measurements from
Netzer \& Trakhtenbrot (2007), kindly provided by B.~Trakhtenbrot,
2007, private communication. For each source we determined the value
of \lledd\ following the procedure outlined in
\S~\ref{sec_properties}. Following the arguments made in
\S~\ref{sec_GE_corr}, the non-simultaneous \xray\ and optical
observations are not expected to affect our analysis considerably (see
also \S\,3 of Steffen \et 2006).

\begin{figure*}
\centering
\plotone{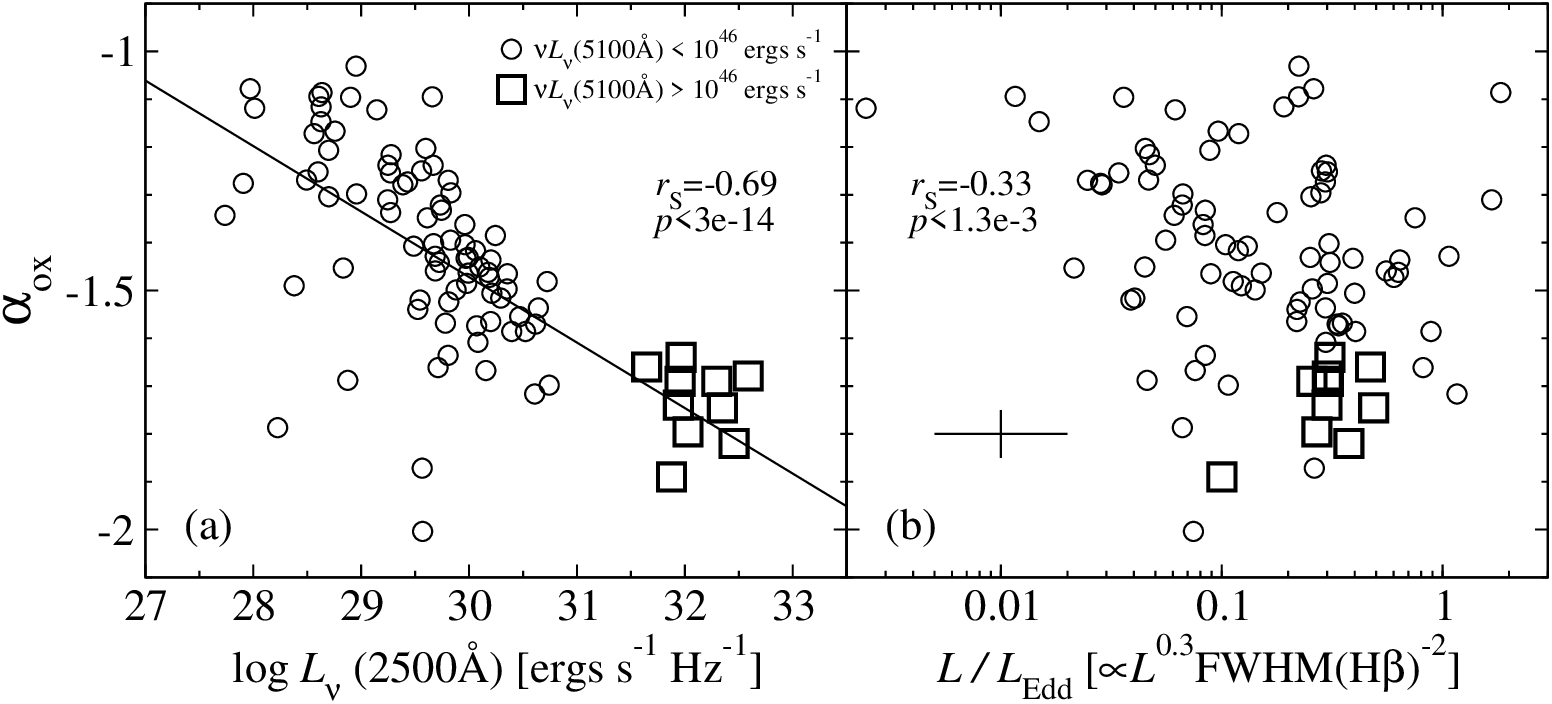}
\caption{Correlations between \aox\ and ({\it a}) $L_{\nu}
  (2500\,\mbox{\AA})$ and ({\it b}) \lledd. Symbols are similar to
  those in Fig.~\ref{fig_GE}, although the low--moderate luminosity
  sources were obtained from Steffen \et (2006). The solid line marks
  the best-fit relationship between \aox\ and $L_{\nu}
  (2500\,\mbox{\AA})$ from Just \et (2007), and the cross marks the
  typical uncertainties on \aox\ and \lledd. Spearman-rank correlation
  coefficients and chance probabilities are indicated in each panel.}
\label{fig_aox}
\end{figure*}

We find that \aox\ is significantly correlated with both $L_{\nu}
(2500\,\mbox{\AA})$ and \lledd; however, the correlation with
luminosity is considerably tighter (see Fig.~\ref{fig_aox}).
Even when a constant bolometric correction factor of 7 is used to
determine \lledd\ values (as performed in \S\,\ref{sec_degeneracy}),
the \hbox{\aox-$L_{\nu} (2500\,\mbox{\AA})$} correlation is
considerably tighter than the \hbox{\aox-\lledd} correlation, although
the significance of the latter correlation has increased (the chance
probability has dropped from \hbox{$p=1.3 \times 10^{-3}$} to
\hbox{$p=1.3 \times 10^{-4}$}).
By investigating this further, we find that the \aox-\lledd\
correlation disappears when the \hbox{\aox-$L_{\nu}
  (2500\,\mbox{\AA})$} correlation is taken into account; i.e., the
difference between the observed \aox\ and the expected \aox\ [based on
the $L_{\nu} (2500\,\mbox{\AA})$ values of the sources and the most
recent \hbox{\aox-$L_{\nu} (2500\,\mbox{\AA})$} relation from Just \et
2007] is not correlated with \lledd.
In another test, we divided the data into sources that have \lledd\
values either above or below the median value of \lledd$=0.22$.
In both sub-samples we detect a strong \hbox{\aox-$L_{\nu}
  (2500\,\mbox{\AA})$} correlation (and no \hbox{\aox-\lledd}
correlation). These tests suggest that the \hbox{\aox-$L_{\nu}
  (2500\,\mbox{\AA})$} correlation is persistent and that \aox\ does
not depend primarily on \lledd.
In this scenario, the (relatively weak) \hbox{\aox-\lledd} correlation
may be a consequence of the inherent dependence of \lledd\ on $L_{\nu}
(2500\,\mbox{\AA})$, and the relative weakness of this correlation
compared with the strong \hbox{\aox-$L_{\nu} (2500\,\mbox{\AA})$}
correlation may be due to the additional uncertainty introduced by
determining \lledd\ (see, for example, the typical error bars in
Fig.~\ref{fig_aox}). Subsequently, if \aox-$L_{\nu}
(2500\,\mbox{\AA})$ is the fundamental correlation, then the inclusion
of the FWHM(\hb) parameter to produce \lledd\ adds substantial scatter
resulting in a rather weak \aox-\lledd\ correlation. It is also of
interest to point out that we find no correlation between \aox\ and
$\Gamma$ (that depends on \lledd) in our core sample of 35 sources
discussed in \S~\ref{sec_GE_corr}.

We also checked whether the \aox-$L_{\nu} (2500\,\mbox{\AA})$
correlation is induced by luminosity-dependent obscuration (e.g.,
Lawrence 1991; Gaskell \et 2004). In a first test, we found a strong
correlation between \aox\ and $L_{\nu} (5100\,\mbox{\AA})$ for our
sample, with similar coefficients as in the \aox-$L_{\nu}
(2500\,\mbox{\AA})$ correlation. We note that the fluxes at 5100\,\AA\
were obtained independently from the fluxes at 2500\,\AA, i.e., no
extrapolations have been made between the flux densities in these two
wavelengths. In a second test, we redefined \aox\ by changing the
optical--UV continuum threshold from 2500\,\AA\ to 5100\,\AA; the
correlations between this modified \aox\ and both $L_{\nu}
(5100\,\mbox{\AA})$ and $L_{\nu} (2500\,\mbox{\AA})$ for our sample
returned similar correlation coefficients as in the \aox-$L_{\nu}
(2500\,\mbox{\AA})$ case.
We conclude that, within the limits of our sample, it is unlikely that
the \aox-$L_{\nu} (2500\,\mbox{\AA})$ correlation is induced by
potential reddening effects.

The fact that $L_{\nu} (2500\,\mbox{\AA})$ appears as the underlying
parameter determining the slope of the \hbox{optical--\xray} SED
instead of \lledd\ is puzzling, since \lledd\ indicates the relative
AGN power and significantly affects the SED of the disk.
One possibility is that our empirical expression for \lledd\ (see S06)
is inaccurate or more complicated. This, however, is not supported by
the fact that $\Gamma$ {\em does} depend on \lledd\ as defined here
and not on $\nu L_{\nu} (5100\,\mbox{\AA})$ (see
\S\,\ref{sec_introduction} and \S\,\ref{sec_degeneracy}; S06). We have
also considered the possibility that the \hbox{\aox-$L_{\nu}
  (2500\,\mbox{\AA})$} correlation is induced by selection effects
(e.g., Green \et 2006), but this is highly unlikely given the fact
that the correlation spans over five orders of magnitude in luminosity
(see, e.g., Just \et 2007). Indeed, future correlations between \aox,
$L_{\nu} (2500\,\mbox{\AA})$, and \lledd\ should involve \lledd\
determinations for \xray-selected sources (see \S\,\ref{sec_BHgrowth}
below).

An alternative approach to interpreting the non-dependence of \aox\ on
\lledd\ stems from the recent results of Vasudevan \& Fabian
(2007). These authors confirm the strong dependence of \aox\ on
$L_{\nu} (2500\,\mbox{\AA})$ in their AGN sample, but they do not find
an \aox-\lledd\ correlation. Utilizing far-UV, as well as optical,
near-UV, and \xray\ spectra for a sample of AGN with known \mbh\ and
\lledd, they argue that the reason for the non-dependence of \aox\ on
\lledd\ lies in the choice of the optical--UV and \xray\ continuum
thresholds (i.e., 2500\,\AA\ and 2\,keV) for the \aox\
definition. While they find dramatic differences between the
optical--\xray\ SED of sources with different \lledd\ values, the
ratio between the flux densities at 2500\,\AA\ and 2\,keV remains
almost constant as \lledd\ changes. The most pronounced differences
between the optical--\xray\ SEDs of different sources are expected to
be concentrated in the $\approx0.01-0.1$\,keV spectral region (i.e.,
the `big blue bump'), which cannot be traced effectively using the
current optical--UV and \xray\ thresholds of \aox. In this context,
the \aox\ parameter, given its current definition, while providing
useful information on the $L_{\rm UV}$-$L_{\rm X}$ connection, cannot
be used as an accretion-rate indicator.
By inspection of Fig.\,13 of Vasudevan \& Fabian (2007), one may
expect a strong dependence of \aox\ on \lledd\ if, for instance, the
optical--UV threshold is shifted from 2500\,\AA\ to as close as
possible to $\sim250$\,\AA, although this is observationally
challenging and may not be practical.
When using modified definitions for \aox, there also remains the
question of which correlation, \aox-\lledd\ or \aox-$L_{\rm UV}$, is
stronger than the other. In general, the choice of optical--UV and
\xray\ thresholds should likely depend upon the scientific question of
interest.

\subsection{The AGN Hard-X-ray Spectrum as a Probe for BH Growth}
\label{sec_BHgrowth}

In \S\,\ref{sec_GE_corr} we have shown that $\Gamma$ may serve as an
\lledd\ indicator in unabsorbed, moderate--high luminosity RQ AGNs
with an acceptable mean uncertainty level on \lledd\ of a factor of
\ltsim3.
Therefore, in principle, given a high-quality \xray\ spectrum in the
\gtsim2\,keV rest-frame band of a RQ AGN, one can empirically estimate
\lledd\ and \mbh\ for the source. Accurate measurements of $\Gamma$
and $L_{\rm X}$ (in the 2--10\,keV rest-frame band, for example) may
provide \lledd\ and \lbol\ using our Eq.\,\ref{eq_lledd_Gamma} and
Eq.\,21 of Marconi \et (2004),\footnote{Vasudevan \& Fabian (2007)
  claim that the bolometric correction factor required to transform
  $L_{\rm 2-10~keV}$ into \lbol\ depends on \lledd, while Marconi \et
  (2004) have used the Vignali \et (2003) \aox-$L_{\nu}
  (2500\,\mbox{\AA})$ correlation (which is not significantly
  different from the most recent correlation of this type given by
  Just \et 2007) to derive bolometric corrections that depend on
  \lbol. Accurate measurements of $\Gamma$ (that provides \lledd) and
  $L_{\rm X}$ will enable comparisons of \lbol\ values obtained from
  these two methods.} respectively, and thus provide an estimate of
\mbh.
We note that the Marconi \et (2004) determination of \lbol\ (whether
from $L_{\rm opt}$ or $L_{\rm X}$) relies, in part, on the very strong
and non-linear $L_{\rm X}$-$L_{\rm UV}$ dependence (see
\S\,\ref{sec_SED}).
The observed (luminosity-dependent) rms errors on that relation are
given in Table\,5 of Steffen \et (2006); these represent typical
deviations of up to $\sim0.15$ from the mean \aox\ [i.e., a factor of
$\sim2.5$ uncertainty on $(f_{\rm
  2\,keV}/f_{2500\mbox{\rm~\scriptsize\AA}})$] for our luminosity
range.
This scatter is inherent in the \lledd-$\Gamma$ correlation as it
reflects uncertainties in determining \lbol\ from $\nu L_{\nu}
(5100\,\mbox{\AA})$ and potential \xray-optical variability.
For example, when \mbh\ values for our core sample of 35 sources are
recovered from \lledd\ using the bolometric corrections for $L_{\rm
  2-10\,keV}$, the ratio between these \mbh\ values and the original
\mbh\ values obtained in \S\,\ref{sec_L5100_Hb} is 1, on average, with
a dispersion of a factor of 2.
In an additional test, we used the $\Gamma$ and $L_{\rm 2-10\,keV}$
values of six nearby sources from our sample to recover their \mbh\
values and compare them with the most recent, direct
(reverberation-mapping based) \mbh\ measurements from Peterson \et
(2004). The six sources, namely PG~0804$+$761, PG~0844$+$349,
PG~0953$+$414, PG~1211$+$143, PG~1307$+$085, and PG~1613$+$658,
comprise $\sim40$\% of all the RQ AGNs with $\nu
L_{\nu}$(5100\AA)$\gtsim 10^{44}$\,ergs\,s$^{-1}$ in the Kaspi \et
(2000) sample (i.e., the PG quasar sample) that have direct \mbh\
measurements. We find that the ratio between the recovered \xray-based
masses and the corresponding \mbh\ measurements from Peterson \et
(2004) is 0.8, on average, with a dispersion of a factor of 1.9; we
also find that the mass recovered for each individual source is
consistent with the measured value, given the uncertainties from
Eq.~\ref{eq_lledd_Gamma} and those from the reverberation-mapping
measurements.

High-quality \xray\ spectra may thus be useful for estimating the
accretion rates and BH masses for moderate--high luminosity type~1, RQ
AGNs and, in particular, for \xray-selected sources. This method may
allow tracing the history of BH growth in the Universe by utilizing
large AGN datasets (see e.g., Brandt \& Hasinger 2005) in which BH
growth cannot be determined effectively using existing methods [i.e.,
reverberation-mapping based methods (aka single-epoch methods), e.g.,
K05, and host-AGN type relations, e.g., Marconi \& Hunt (2003)],
either due to the faintness of the sources in the optical--near-IR
bands, or that the required spectroscopic features are either not
accessible or cannot be modeled reliably. One advantage of the
\xray-based method is the ability to obtain \lledd\ and \mbh\ by
measuring broad spectroscopic properties (i.e., $\Gamma$) as opposed
to detailed spectroscopy for obtaining the width of an emission line
(e.g., \hb\ or \ion{Mg}{2}; this involves, for example, careful
decontamination of \ion{Fe}{2} emission features from the UV--optical
spectra). Moreover, measurements of the hard-\xray\ power-law spectral
slope are typically not limited to specific redshift ranges, such as
those dictated by atmospheric transmission bands (and detector
bandpass) for ground-based spectroscopy. These advantages offer a way
to obtaining many \lledd\ and \mbh\ estimates, economically, in
contrast with more complicated (and redshift-restricted) spectroscopic
measurements for individual sources done with existing
methods. Nevertheless, we caution that the \xray\ measurements of
$\Gamma$ and $L_{\rm X}$ should be done carefully to account for
potential complex absorption and Compton reflection.

The \xray-based method for estimating \lledd\ and \mbh\ may prove to
be even more rewarding in cases where a source is either optically
faint and/or the broad-emission lines are too weak to measure if, for
example, these lines are overwhelmed by host-galaxy continuum (e.g.,
Moran \et 2002; provided the source is not radio loud and that the
redshift can be determined). In particular, this method may be the
best way to determine \lledd\ and \mbh\ directly in obscured (i.e.,
optical type\,2), moderate--high luminosity AGNs. Provided the \xray\
absorption column density is not too high
(\nh\ltsim$10^{23}$\,cm$^{-2}$) and can be modeled accurately, the
intrinsic hard-\xray\ power-law spectrum can, in principle, provide
\lledd\ and \mbh\ as outlined above. Examples of such sources with
high-quality \xray\ spectra are given in e.g., Civano \et (2005),
Mateos \et (2005), and Mainieri \et (2007). Since Compton-thin type~2
AGNs comprise a significant fraction of the AGN population, tracing
the growth of the supermassive BHs in their centers using the
\xray-based method is crucial for testing models of the evolution of
BH growth and accretion luminosity in the Universe in an unbiased way.

\section{Conclusions}
\label{sec_conclusions}

We present \xray\ spectroscopy for five highly luminous RQ AGNs at
$z=1.3-3.2$, with accurate FWHM(\hb) measurements that allow
determinations of their \mbh\ and \lledd\ values. Analysis of the
\xray\ spectra provided measurements of the hard-\xray\ photon index
in the rest-frame \gtsim2\,keV band and \aox. We have combined these
data with the S06 sample of 30 moderate--high luminosity sources with
similar properties, while doubling the number of highly luminous
sources in their sample. Our main goal was to test the S06 claim that
$\Gamma$ can serve as an accretion-rate indicator in RQ AGNs. We have
also tested whether any additional \xray\ properties of our highly
luminous sources depend on \lledd. Our main results are summarized as
follows:

\begin{enumerate}

\item{Our new highly luminous sources with FWHM(\hb) measurements have
    allowed us to break the degeneracy between the dependence of
    $\Gamma$ on FWHM(\hb) and on \lledd, suggesting that the accretion
    rate largely determines the hard-\xray\ spectral slope across four
    orders of magnitude in AGN luminosity (i.e., $44\ltsim \log [\nu
    L_{\nu} (5100~\mbox{\AA})] \ltsim 48$).}

\item{We found a significant correlation between \lledd\ and $\Gamma$
    with a best-fit line of the form $\log (L/L_{\rm Edd}) =
    (0.9\pm0.3) \Gamma - (2.4\pm0.6)$, and an acceptable uncertainty
    of a factor of \ltsim3 on a predicted value of \lledd.}

\item{Utilizing a sample of 91 sources from Steffen \et (2006) and
    this work, we find that \aox\ depends strongly on optical--UV
    luminosity and only weakly on \lledd; the (weak) correlation with
    \lledd\ is probably due to the strong \hbox{$L$-\lledd}
    dependence. We discuss possible explanations for this result
    including the possibility that \aox\ cannot be used as an
    accretion-rate indicator based on its current definition.}
    
\item{We find a significant Compton-reflection feature in two of our
    sources, and the mean relative reflection for seven other sources
    is $R=0.9^{+0.9}_{-0.6}$. By setting rather loose constraints on
    the strengths of \Ka\ emission lines in our highly luminous
    sources, we can neither confirm nor rule out a suggested
    anticorrelation between EW(\Ka) and either luminosity or \lledd;
    the upper limit on the mean rest-frame EW(\Ka) for seven of these
    sources that do not show Compton-reflection features is 105\,eV.}

\item{We have not detected any signature of a soft-excess component in
    any of our highly luminous sources, including two sources at
    $z\sim1.4$ where our rest-frame coverage extends to
    $\sim0.7$\,keV, suggesting that the soft excess does not depend
    strongly on the accretion rate.}

\item{Although one of our highly luminous sources, LBQS\,0109$+$0213,
    exhibits long-term (i.e., on timescales of years) \xray\
    variations, rapid \xray\ variations on timescales of $\sim$1\,hr
    have not been detected in any of our highly luminous (and
    high-\mbh) sources, supporting the idea that \xray\ variability
    timescale depends inversely on \mbh\ and does not depend on
    \lledd.}

\end{enumerate}

The strong correlation between $\Gamma$ and \lledd\ may serve as a
useful probe for tracing the history of BH growth in the universe. It
may provide \lledd\ and \mbh\ estimates for \xray-selected AGNs, with
the possibility of estimating these properties for Compton-thin
type\,2 AGNs for the first time.

\acknowledgments

This work is based on observations obtained with \xmm, an ESA science
mission with instruments and contributions directly funded by ESA
Member States and the USA (NASA). We thank an anonymous referee for a
helpful report that assisted in improving the presentation of this
work. We are also grateful to Franz Bauer, George Chartas, Brandon
Kelly, and Aaron Steffen for useful comments and fruitful
discussions. We gratefully acknowledge the financial support of NASA
grants \hbox{NNG05GP00G} and \hbox{NNX07AE77G} (O.\,S, W.\,N.\,B),
NASA LTSA grant \hbox{NAG5-13035} (O.\,S, W.\,N.\,B), and the Zeff
Fellowship at the Technion (S.\,K). This work is supported by the
Israel Science Foundation grant 232/03.

\end{document}